\documentclass[preprint,aps,prd,nofootinbib,superscriptaddress]{revtex4}
\usepackage{graphicx}

\newcommand{\beq}{\begin{equation}}
\newcommand{\eeq}{\end{equation}}
\newcommand{\beqs}{\begin{eqnarray}}
\newcommand{\eeqs}{\end{eqnarray}}

\def\Tau{T}

\begin{document}

\title{Characterizing Planar Graphs}

\author{Vladimir Gudkov}
\thanks{gudkov@sc.edu}

\affiliation{  Department of Physics and Astronomy,
        University of South Carolina,
           Columbia, SC 29208}

\author{Shmuel Nussinov}
\thanks{nussinov@post.tau.ac.il}

\affiliation{ School of Physics and Astronomy, Tel Aviv University,
Tel Aviv, Israel}

 \affiliation{
Schmid College of Science, Chapman University, Orange, CA 92866}

\author{Zohar Nussinov}
\thanks{zohar@wuphys.wustl.edu}

\affiliation{Department of Physics, Washington University, St.
Louis, Missouri 63160}

\affiliation{Kavli Institute for Theoretical Physics, Santa Barbara, CA 93106}

\date{\today}

\begin{abstract}

 Cataloging planar diagrams using the depth concept is proposed.
\end{abstract}

\maketitle

 \section{Introduction}

 The chararacterization of all graphs is a hard problem. No existing algorithm
 can test, in polynomial number of steps, if \emph{any} pair of graphs, represented
 by $n \times n$ adjacency matrices with $n$ being the number of vertices, are topologically equivalent.

Planar graphs -- namely those graphs which can be drawn on a plane
or a sphere without crossing of any two edges -- are a special,
simple subclass of graphs. Testing for planarity, for equivalence of two
planar diagrams, and finding Hamiltonian paths on a planar graph 
(paths in which each vertex of the graph is visited once and only once)  can all be done in
polynomial time. Also, simple algorithms transcribed into codes
prescribing the construction of any planar graph exist in the
mathematical literature \cite{NogaAlon}.

 Here, we characterize planar graphs using a more ``physical'' approach based on the
 concept of depth to be defined below.

 Interest in planar Feynman diagrams in the high energy community was motivated by 't-Hooft's
 realization \cite{'tHooft:1974pg}  that in the large $N$ limit of $SU(N)$ non-abelian gauge theories,
 planar diagrams dominate. More recently, these have been the focus of interest in the condensed 
 matter community.  \cite{sung-sik}
 Diagrams which can be drawn without crossings only on a surface
 of genus $g$ are $O(N^{-2g})$ in a $1/N$ expansion. The fact that the total number of planar graphs
  (or finite genus $g$) increases only as $c^n$ ( rather than as $n!$) was found \cite{Koplik:1977pf}
  by using the mathematical literature (mainly by Tutte) on triangulations or by directly solving
  a zero dimensional field theory \cite{Brezin:1977sv}.
More recently, planar graphs were used in slightly more intricate
limits
 with fermions in appropriately chosen representations in addition to the gauge bosons in
 \cite{Armoni:2003fb}.

 Explicit summation of all planar diagrams and exact solutions have been achieved only in the $1 + 1$
  dimensional case by 't-Hooft \cite{'tHooft:1974hx} using the fact that only iterated rainbow diagrams
  contribute to the quark propagator. Further characterizations of planar diagrams may help actual computations.

 Mesoscopic devices or printed circuits with predominantly intra (single) plane connections,
 and highway systems, are also almost planar and  characterization of these by depth may be helpful.

 \section{Generalities and the Depth in Planar Graphs}

  The connectivity, distance, diameter, valency, and other basics can be defined for any graph.
  The concept of depth defined next is, however, unique to planar graphs. For
  simplicity,   we consider planar $\phi^3$ diagrams, namely, allow only valency $v=3$ vertices.
 Lets start with a diagram in coordinate space with no external lines drawn on a sphere.
  We allow one vertex to have an arbitrary valency, move it to the north pole, and stereographically
   project on the plane. The special point will be at infinity and the lines from it become the $k$
    external lines or particles coming from (going to) infinity.

In general, graphs with self energy and tadpole insertions
correspond to several different planar drawings.
 To simplify, we will assume that our graphs are tadpole free and without self energies.

The region around the special vertex (i.e.,  the point of infinity) is
the ``exterior'' of the graph. The various adjacent ``faces'' of the
graph are divided into classes according to ``depth.'' The depth of
a face is the minimal number of edges that need to be cut in order
to reach it from the exterior.

 In a ``dramatic'' description we view the edges in the planar drawing as the walls
  of a city besieged by outside enemies. Each edge is the boundary between two faces
  of depths $D$ and $D'$.  The corresponding wall, or edge, is assigned a height
  $h$
  \begin{equation}\label{e1}
    h=\max \{D,D'\}= \min \{D,D'\} +1 \hspace{1.0cm}  \mbox{if} \hspace{0.5cm} D\neq D'
  \end{equation}
or
\begin{equation}\label{e2}
    h=D \hspace{1cm} \mbox{if} \hspace{1cm} D=D'.
\end{equation}

At each stage, the enemy -- or a rising sea encircling our island
city -- scales (rises above)
 all walls of given height $h$.  Next, it rises to height $h+1$ , etc. Thus, at each
 stage,
 the ocean rises by one extra unit and floods faces one unit deeper.

\section{Going from One Depth to the Next}

 The gradual flooding process allows  peeling off successive layers of the besieged
   island  or eventually of each of the separate islands  previously generated.
   This allows a recursive definition of the complete graph.

\begin{figure}
\includegraphics[scale=0.6]{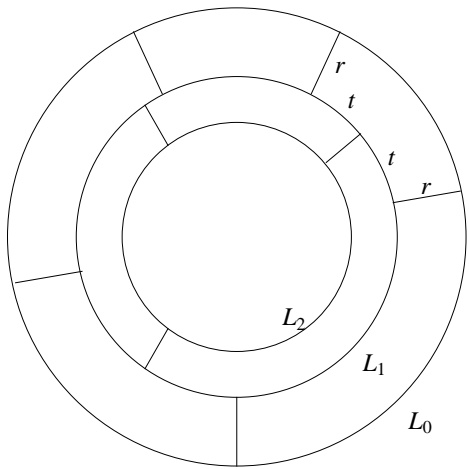}
 \caption{\label{f1}
 A series of concentric loops $L_0$, $L_1$, and $L_2$ in our graph are connected by radial ``spokes''
  denoted by $r$. Each of the loops is made of a consecutive series of ``tangential'' edges: 5 on $L_0$, 8  on
  $L_1$, and 3 on $L_2$ in this particular case.}
 \end{figure}

 Edges of type (\ref{e1}) or (\ref{e2}) above, are termed ``tangential'' or ``radial''
 respectively,
  a nomenclature suggested by the special graph with concentric circles threaded by radial ``spokes''
   emanating from the center (see Fig. \ref{f1}).

    The tangential edges contribute to the  defense of our island from the steadily
     rising sea, whereas radial edges, needed for the connectivity of the graph, do not help this defense at all.
     Thus the two faces flanking a radial edge have equal depths and get flooded simultaneously. On the other
     hand, any tangential edge of type (1) separates two faces of depths differing by one unit.
     Hence, once the lower depth face is flooded, the tangential wall in question still protects the ``inner,''
     deeper face. A series of contiguous tangential edges separates an ``outer'' region of depth $D$ from an ``inner''
      region of depth $D+1$. At some point, the depth $D$ faces will all be flooded by the ``ocean'' which
      constitutes one continuous connected body of ``water.''
      Hence, the chain of contiguous tangential edges cannot be terminated at a vertex. In such a case, the ocean would
      have ``flanked'' this chain on both sides, and the chain cannot separate two regions. The boundary of any
      region has no boundary and a continuous chain of tangential line must form a closed ``circular'' loop.

In general, the complete set of tangential lines defending regions
of a given depth $D$ form several disconnected loops, each enclosing
a separate ``island'' which is a smaller subgraph of the original
graph. An important simplifying feature is that these inner planar
subgraphs -- being physically separated from the outside region, and
from each other -- can be independently prescribed. This is the key
to our recursive definition. The main steps involved amount to
repeated constructions either of actual tree subgraphs or using
``tree like'' combinatorics.

\begin{figure}
 \includegraphics[scale=0.6]{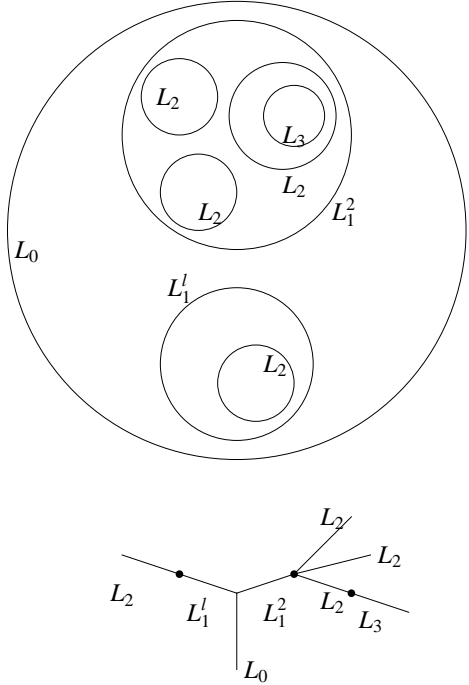}
 \caption{\label{f2}
The hierarchy of inclusions of the circular loops in a planar graph.
$L_0$ includes the two separate loops $L_1^1$ and $L_1^2$. $L_1^1$
includes one circular loop $L_2$,  and $L_1^2$ includes three $L_2$
loops, of which one includes one $L_3$ loop.  The resulting short
tree embodying this inclusion hierarchy is shown under the big loop.
 Note that all radial and tangential edges are omitted here.}
 \end{figure}

 Consider first the ``overall'' depth structure of the graph as manifest by drawing \emph{all}
 the closed loops encircling the islands arising at all the stages of the flooding process.
 These consist of the large overall boundary encircling the complete graph of depth $D=1$.
 Inside this loop, we have one or up to $b_1$ separate loops each encircling a subgraph of depth $D=2$.
 Inside each of the above ``first generation'' loops, we have one or several loops, all together adding up to $b_2$
  ``second generation'' loops, etc ( see Fig. \ref{f2}).

 This hierarchy of inclusions is tree-like. Thus, consider a real physical tree growing vertically
 and dividing into branches, each branch next divides into branches of the next generation, etc.
 By cutting the full tree at each of the levels and projecting all cuts, we generate exactly the original hierarchical
  set of loops made up of all the tangential edges. The equivalent -- in this case, a rather simple ``tree of loops''
  -- is illustrated at the bottom of Fig. \ref{f2}.

 We still have to construct the remaining set of ``radial'' edges in the region between an ``external'' loop
 around a region of depth $D$ and in general several smaller ``inner'' loops each encircling a subgraph of depth $D+1$.

\begin{figure}
 \includegraphics[scale=0.6]{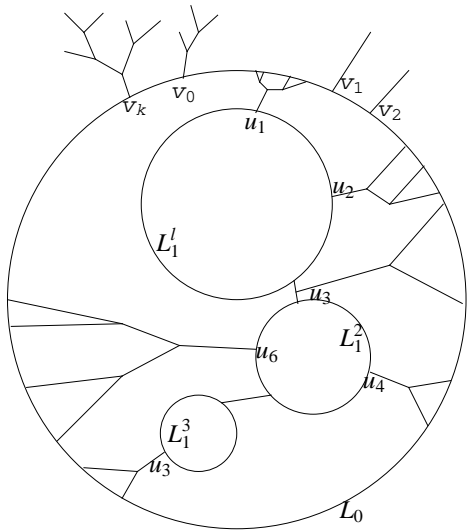}
 \caption{\label{f4}
A Tree of the three $L_1$ loops and the actual $\phi^3$ trees with
tips ending on the external $L_0$.}
 \end{figure}

 This construction repeats itself at each depth or generation. We start with the complete original diagram but will
 keep in mind also the analog repeats in later generations. The
$E^0$ external lines can merge into trees $T^0_1$, $T^0_2$, $ \ldots
T^0_i \ldots T^0_k$, with $T^0_i$ having $t^0_i$ tips, so that
$\Sigma t^0_i= E$. The $k$ trees are ``rooted'' along the overall
encircling loop $L_0$ at $k$ vertices as indicated in Fig.
\ref{f4} bellow. For concreteness, we define
one such vertex as $V_0$ and number the rest in a clockwise
direction. These external vertices then carry the complete
information pertinent for all subsequent steps performed inside as
described below, which is inherited from the outside.

\begin{figure}
\includegraphics[scale=0.6]{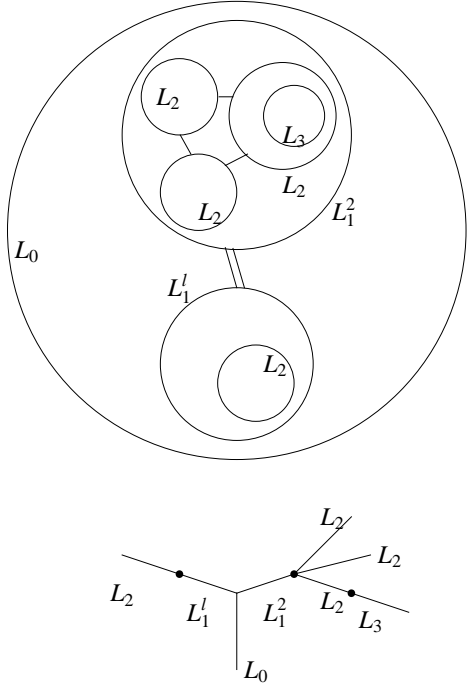}
 \caption{\label{f3}
Illustrating why we cannot connect two circular loops $L_1^1$ and
$L_1^2$ of the same depth (or the same generation) by two lines
since we then will join their interiors to one big overall loop made
of the two loops - in which case, we will have only one $L_1$ loop
rather than two, as we assumed to be the case. Also, connecting the
three $L_2$ loops by three lines will generate one $L_2$ type
loop inside the $L_1^2$ loop -- rather than three, as was assumed to
be the case.}
 \end{figure}

Consider next the $L_1$ loops encircling islands of depth $D= 2$,
which are exposed after  the outside big loop  eroded. We can connect
those to each other via ``radial'' edges. We should not connect two
loops by more than one line. Otherwise, rather than having two
different circular loops, they will be joined to one bigger circular
loop. More generally, we should not form bigger new  $L_2$ type
loops in the next generation using the new lines and portions of
the above $L_2$ loops  (see e.g. Fig. \ref{f3}).
Otherwise, this bigger loop, rather then the individual smaller
loops, will be exposed after the first stage. Thus, we can only
connect the loops by single lines into one overall ``tree'' (whose
vertices are the above $L_1$ loops) or to several separate such
trees (see Fig. \ref{f4}).
 Apart from partitioning the $L_1$ loops into the subsets
forming the vertices of these trees, we have again just tree
formation.

Next, we  add radial edges joining the above trees of loops to the
outer big overall loop.
 Some such edges are needed to ensure connectivity. To avoid modifying the first generation circular
 loop(s),
  the connection to the ``outer frame'' should be only via trees rooted on the inner loops with tips on the outer loop.
These trees are similar to what we assumed above for the external
lines forming trees $T^0_i$ that are attached from the outside
to the external loop. Indeed, by construction, the trees offer no further
defense when the outside big loop is stripped off, conforming to the
definition of radial edges of which these trees are made. Being
directly parts of the graph, the assumed threefold valency implies
that these are simple binary trees. There is still the question of
how to place the tree tips from the inside between the tree roots
from the outside without violating the required planarity.

 We start first with the simple case with just \emph{one} inner loop.
There are special points on both the external and internal loops,
namely, the roots $v_i$ of the external particle trees  and
the roots of the internal trees $u_i$. Each ``first
generation'' tree, $\tau^1_j$ maps, via its tips, the $j$-th point
on the internal loop to an interval $I_j$ on the outer loop where
all the tips reside. Each interval $I_j$ is flanked by two external
points $r^0_l(j)$ and $r{^0}_l'(j)$ with $l'-l\ge 0$ being
 the number of external points in this interval. In the case when for
 all  $j$: $l(j)>l(j')$ if $j>j'$, we have no line crossings of edges of the
  different trees and planarity can be maintained.

\begin{figure}
\includegraphics[scale=0.6]{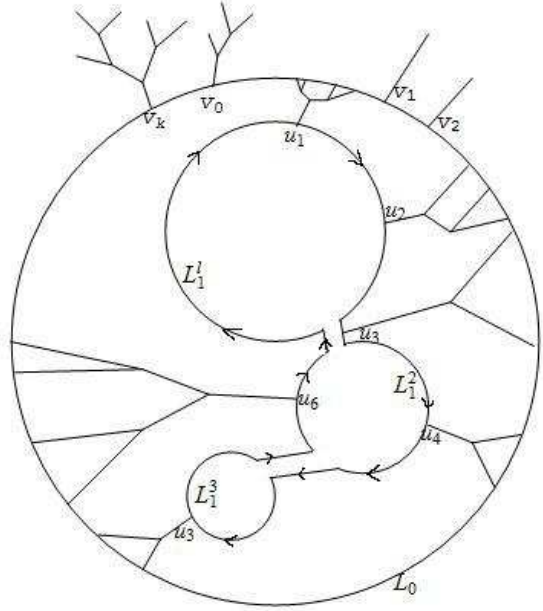}
 \caption{\label{f5}
Connecting the three loops in the previous graph with fictitious
double lines so as to allow a cyclic numbering of all the vertices
on these loops (and/or on the connecting lines).}
 \end{figure}

 By assumption all the $L_1$ first generation loops are connected so as to form a single tree of loops.
 We will denote such ``trees of loops'' by $T$'s and the ``true'' $\phi^3$ binary trees connecting
 them to the external loop by $\tau$'s.
Next -- just for the purpose of ordering the vertices $u_i$ -- we imagine doubling the lines connecting  the
 tree of loops $T$ (see Fig. \ref{f5}). This induces a common clockwise
ordering\footnote{
 This is the case for graphs drawn on orientable manifold including higher Genus surfaces.}
  of all the $u_i$ vertices and the trees emanating from the various loops and
possibly also from the connecting lines.

 The $L_1$ first generation loops can form several  (say $F$) disconnected ``trees of loops'' $\Tau_{(f)}$, $f=1,\ldots
 F$,
 with the $k_f$ roots of the ``real'' binary trees residing on the loops (or connecting lines); and each of the $F$ trees
 of loops can separately be clock-wise ordered as above.
We then have $\sum_ {f=1}^{F} k_{f}= N^1(\Tau)$ binary $\tau$
trees rooted on the first generation loops or on the lines between
those   and connected to the external
overall loop.

In addition we can have some number $N^1(\tau)$ of ``hanging'' binary
trees attached only to the external loop. The last stage of defining
the recursive process of going from depth $D$ to depth $D+1$
involves placing the tips of the above $ N^1=N^1(T) +N^1(\tau)$,
first ($(D+1)$-th in general) generation trees relative to roots of
exterior (in general  $D$-th generation) trees on the surrounding
big loop. This involves further combinatoric which we discuss in the
next section.

\section{The "Tree" of the Nesting Hierarchy of Trees}

 Connecting the $n^1_j$ tips of the first ( $(D+1)$-th) generation  $j= 1,2,\ldots N$ trees  to the external
  ($D$-th generation) loop again involves ``tree like'' constructions.

\begin{figure}
\includegraphics[scale=0.6]{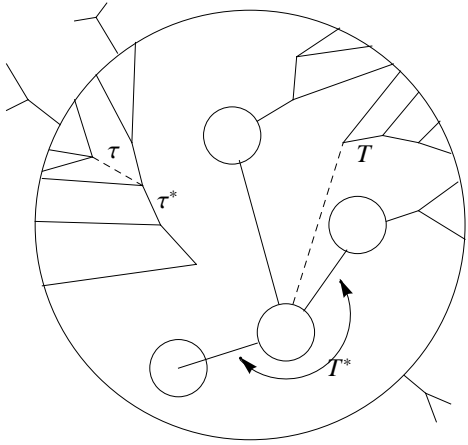}
 \caption{\label{f6}
The pattern of nesting indicated by the fictitious broken lines
connecting the root of the nested tree to the branching point in the
higher echelon nesting tree.}
 \end{figure}

Any of these trees can be ``nested'' within consecutive branches of
another tree, and other trees can be nested between pairs of its
branches.  The ``lowest''in this hierarchy are trees which harbor no
other trees. The tips of any of these ``non-harboring'' trees, $\tau$,
define a continuous single interval $I(\tau)$ on the exterior
encircling loop on which all its tips reside. (This interval must
contain at least one root of the external trees or, more generally,
a root of a tree from a previous generation. Otherwise, the tree $\tau$
hangs from one tangential edge making a ``mass'' insertion which was
excluded by assumption).  One notch higher in the nestling hierarchy
are $\tau^*$ trees which nestle lowest echelon of trees, $\tau$'s . Two
tips of $\tau^*$ should land on the external ring on the two sides of
the interval $I(\tau)$ associated with the nestled tree $\tau$'s.
Recall that the interval $I(\tau)$ contains
all the tips of the tree $\tau$. This
should be the case for all intervals $I(\tau)$ corresponding to all
trees  $\tau$ nestled within $\tau^*$. All these intervals together with the tips
of $\tau^*$ then cover a larger continuous interval $I (\tau^*)$, the
overall domain of influence of the harboring tree $\tau^*$. In turn,
several such $\tau^*$'s and possibly an additional first echelon of
trees $\tau$'s nestle within yet a higher (in nestling hierarchy) tree
$\tau^{**}$, etc. This can keep on going until all the radial edges
composing trees connecting the $(D+1)$-th generation loops to the
encircling single $D$-th generation loop are exhausted. To specify
fully
 the nestling, we need not only to know  which tree $\tau^*$ is a
given $\tau$ tree nested in, but  also to specify the two branches
between which  the lower echelon is nested, as is the case here with
a $\tau$ tree. As indicated in Fig. \ref{f6}, it is specified by an
extra ``pointer edge'' (shown as a broken line) connecting the
branching point of the $\tau^*$ tree and the root of the tree $\tau$
nested therein.

Thus, to specify fully  the set-up bridging between two consecutive
depths, we formally need an extra ``field'' $\psi $, whose propagator
is the above broken line (as opposed to the propagator of the main
field $\phi$ denoted by the full edges in the diagram) and also, in
addition to the true $\phi^3$ vertices of the original graph, some
$\phi^3\psi$ vertices to allow the above pointers.

 Finally, in this nestling process, we cannot treat the hanging $\tau$ trees
 and the loop ($T$) trees the same way.

When nested within a higher echelon tree, the \emph{whole} $T$
tree of loops along with the many ordinary $(\phi^3)$ $\tau$-type
trees spawned off from it are both treated as one nested unit.
Still, we need to maintain  the correct ordering on the external loop the  of the $\tau$ trees
spawned from the loops of the  $T$ trees, and the correct cyclic
ordering relative to the external loops above has to be maintained.

 Further, when nesting trees within a higher echelon trees, the $T^*$ and $\tau ^*$ trees
  are also clearly different. The point is that each $T^*$ tree ``spawns'' several regular $T$ trees,  and we need to specify in which $T^*$ tree (we are using the broken line
   indicators for
   these specific junctions) is the $T$ tree is nested.

The process of defining successively broader domains of influence on
the external loop, each containing contiguous domains of influence
of lower hierarchy trees, is clearly isomorphic to building a
general ``nestling tree'' and even pictorially is similar to the
basic tree made of tangential loops of higher depth (Fig. \ref{f3}).
There, the largest, most inclusive, outermost loop  is indeed
the root of the abstract inclusion tree. Here, there is some ambiguity as
to which tree provides the root of the ``nestling hierarchy tree'':
we can sometimes declare that  tree $A$ is nested within $B$ or,
reversing the point of view, that $B$ is nested within $A$.
 We avoid this ambiguity by using the "origin," namely, the first vertex $v_0$ on the external loop.
 It will induce the "origins" on each of the internal loops, namely the point on each loop closest to it in a clock wise sense.
   We declare the overall nestling tree -- the one highest in the hierarchy of nesting (encountered
    in any of the depth=$D$ $\longrightarrow$ depth= $(D+1)$ transition) --
    to be the one whose domain of influence includes the origin on the encircling loop.

\section{Summary and Further Comments.}

 The above recursive definition of the general planar graph by means of the concept of depth is clearly \emph{not}
  the simplest or the most concise  possible. In such terms, it is significantly inferior to the concise formal
  mathematical construction. We hope, however, that this particular method, being based on a heuristic ``physical''
   notion of depth, may be useful in some physical applications.
 Specifically, in computing planar Feynman diagrams, it is interesting to ask if the tangential loops and
 radial trees can offer a more natural way for routing the momenta.
(The question as to the optimal ``choice of loops'' occurs already
for electrical networks.) Thus, rather than using the individual
elementary faces of the planar graph for the loops, we can use  the
``circular'' loops made of ``tangential'' edges. All the external
moments (or currents) can then be made to flow only in the external
overall loop. It is often the case that the sum of one loop diagrams
with any number of external lines defines a useful effective action.
By envisioning that the planar diagrams are in configuration space we may have here
effective actions on different scales.
Then, the effective actions of larger and larger circular loops may
correspond to steps in a renormalization group that begin with short
distances and then build up the exact (in the planar approximation)
infrared physics.

 The fact that all the combinatorics involved in our construction
  of the diagrams was essentially of repeated construction of trees suggests, at
   different levels of abstractions that
there may be some reasonably simple semi classical description.

\section{Acknowledgements}

 We are indebted to Noga Alon for very useful discussions of the above subject and other graph connected issues.

%\begin{thebiblography}{99}

\end{document}